\documentclass[twocolumn,showpacs,preprintnumbers,amsmath,amssymb,nofootinbib,superscriptaddress]{revtex4}
\usepackage{hyperref}
\usepackage{graphicx}
\usepackage{color}

\newcommand{\be}{\begin{equation}}  
\newcommand{\ee}{\end{equation}}  
\newcommand{\bear}{\begin{eqnarray}}  
\newcommand{\eear}{\end{eqnarray}}  
\newcommand{\ba}{\begin{array}}  
\newcommand{\ea}{\end{array}}

\newskip\humongous \humongous=0pt plus 1000pt minus 1000pt

\newif\ifdtup

\def\oldreffmt#1{\rlap{[#1]} \hbox to 2\parindent{}}

\def\figfmt#1{\rlap{Figure {#1}} \hbox to 1in{}}  
  
\def\ie{\hbox{\it i.e.}{}}	  
\def\eg{\hbox{\it e.g.}{}}

\def\beq{\begin{equation}}  
\def\eeq{\end{equation}}  
\def\bea{\begin{eqnarray}}  
\def\eea{\end{eqnarray}}

\def\bq{\begin{quote}}  
\def\eq{\end{quote}}

\relax  

\newdimen\tdim  
\tdim=\unitlength  
\def\bar{\overline}

\begin{document}

{\title{Is the  Higgs Boson Composed of Neutrinos?}

\author{Jens Krog}
\email{krog@cp3-origins.net}
\affiliation{CP3-Origins, University of Southern Denmark\\
Campusvej 55, 5230 Odense M, Denmark\\$ $\\}

\author{Christopher T. Hill}
\email{hill@fnal.gov}
\affiliation{Fermi National Accelerator Laboratory\\
P.O. Box 500, Batavia, Illinois 60510, USA\\$ $}

\date{\today}

\begin{abstract}
We show that conventional Higgs compositeness conditions can be achieved by the running of large
Higgs-Yukawa couplings involving right-handed neutrinos that become active at $\sim 10^{13}- 10^{14}$~GeV.
Together with a somewhat enhanced quartic coupling,
arising by a Higgs portal interaction to a dark matter sector, we can
obtain a Higgs boson composed of neutrinos.  This is a
"next-to-minimal" dynamical electroweak symmetry breaking scheme.

\end{abstract}

\maketitle

\section{Introduction}
Many years ago it was proposed that the top quark Higgs-Yukawa (HY) coupling,   $y_t$, might
be large and governed by a quasi-infrared-fixed point behavior of the
renormalization group \cite{PR,CTH1}. This implied, using the minimal ingredients of the
Standard Model, a top quark mass of order $220 - 240$ GeV for the case of a Landau pole in
$y_{t}$ at a scale, $\Lambda$, of order the GUT to Planck scale. 
In light of the observed 173 GeV top quark mass, the fixed point
prediction is seen to be within 25\% of experiment. This suggests that small corrections
from new physics might bring the prediction into a more precise concordance with
experiment.   

One of the main interpretations of the quasi-infrared fixed point 
was the compositeness of the Higgs boson. In its simplest form, the
Higgs boson was considered to be a bound state containing a top and anti-top quark
\cite{Nambu,Yamawaki,BHL,other}. This was amenable to a treatment in a large-$N_c$ Nambu--Jona-Lasinio
model \cite{NJL}, defined by a  4-fermion interaction at a scale $\Lambda$, with a 
a large coupling constant, and a strong attractive $0^+$ channel.  The theory requires drastic
fine-tuning of quadratic loop contributions, which is equivalent to a
fine-tuning of the scale-invariant NJL  coupling constant (and
may require a novel insight into scale symmetry, such as \cite{bardeen2}).
By tuning the NJL coupling close to criticality, the Higgs boson mass
becomes small, creating an infrared hierarchy between the compositeness
scale, $\Lambda$, and the electroweak scale embodied in $m_h$.
Tuning the coupling slightly supercritical yields a vacuum instability 
and the Higgs boson acquires its VEV.

Once the infrared hierarchy has been tuned,
the remaining structure of the theory is controlled by  renormalization group (RG) running of 
couplings \cite{BHL}.
The RG treatment indicates that a $\bar{t}t$ composite Higgs boson requires
(i) a Landau pole at scale $\Lambda$ in the running top HY coupling constant, $y_t(\mu)$,  (ii) the Higgs-quartic coupling $\lambda_H$ must
also have a Landau pole, and (iii) compositeness conditions must be met, such as
$\lambda_H(\mu)/g_t^4(\mu)\rightarrow 0$  and $\lambda_H(\mu)/g_t^2(\mu)\rightarrow $(constant) as
$\mu \rightarrow \Lambda$,  \cite{BHL}.
This predicts a Higgs boson mass of order $\sim 250$ GeV with a heavy top quark of order $\sim 220$ GeV, 
predictions that come within a factor of $2$ of reality.

While the $\bar{t}t$ minimal composite Higgs model is ruled out, it remains
of interest to ask, ``can we rescue an NJL--RG  composite Higgs boson scenario with new physics?''
and if so, ``what are the minimal requirements of new physics needed to maintain a composite Higgs
boson scenario?"  
In the present paper we address this issue
and revisit a composite Higgs boson model based upon  
an attractive idea of S. P Martin,  \cite{Martin}
(this has also been considered in a SUSY context by Leontaris, Lola and Ross \cite{Ross}).  Martin
pointed out that the top quark HY is sensitive to right-handed neutrinos, $\nu^i_R$, that become active
in loops above the large Majorana mass scale, $M$.  The right-handed neutrinos
are assumed to have HY couplings, $y_\nu^i \geq {\cal{O}}(1)$,
 and also have  a Majorana mass of order $M\sim 10^{13}$ GeV, thus leading to the
neutrino seesaw model at low energies \cite{seesaw}.  Turning on the neutrino loops will generally
pull a large $y_t(m_{t})$ to a Landau pole at a scale of order $\Lambda \sim 10^{15} - 10^{19}$ GeV,
and  the large top quark mass becomes intertwined with neutrino physics above $M$.
The strong dynamics that forms the boundstate Higgs boson for us
is the dominant large coupling,  $y_\nu^i $.

Martin's model preserved some of
the features of the $\bar{t}t$ composite Higgs model, but extends
 the Higgs compositeness structure to become an entanglement of neutrinos and top quark.
Martin's model considered one large neutrino HY interaction, while we will
presently consider $N_f=3$ right-handed neutrinos with degenerate  HY couplings
to the corresponding left-handed doublets, $y_\nu$.  Hence, our present model
becomes a large-$N_f$ fermion bubble NJL model as we approach the compositeness scale $\Lambda$.
The $y^i_\nu$ become active above the scale 
$M$ and are chosen to be large enough to have Landau poles at the scale $\Lambda$.

The top quark HY coupling is pulled up by the large  $y^i_\nu$ to a Landau pole, but we 
find that the ratio of $y_t(\mu)/y_\nu(\mu)\rightarrow $~(constant) as $\mu\rightarrow \Lambda$.
This implies that the top quark  couples to the
dynamics that forms the Higgs boson at the scale $\Lambda$, which we
treat as a Nambu--Jona-Lasinio model, but this is merely a comparatively weak
 extension of the dynamics to give mass to the top (and presumeably all other
quarks and leptons). The Higgs doublet (Higgs scalar) in our scheme is primarily
composed of $\sum_i\bar{L}_{iL}\nu_R^i$, ($\sum_i\bar{\nu}_{i}\nu^i$),  where  $L_{iL} = (\nu_i, \ell_i)_L$
is the left-handed lepton doublet, and summing over $N_f=3$ generations. 
The top quark HY coupling is then only a spectator to this physics. 

A second and important demand of an NJL-composite Higgs model is the behavior of the running of the
Higgs quartic coupling, $\lambda_H$.  It is show in ref.\cite{BHL} that $\lambda_H$ will
have a Landau pole at the scale $\Lambda$, but  $\lambda_H(\mu)/y_{t}^4(\mu)\rightarrow 0$
and  $\lambda_H(\mu)/y_{t}^2(\mu)\rightarrow $~(constant) as $\mu \rightarrow \Lambda$. 
Engineering this is more challenging issue than that of the 
Landau poles in the Yukawa couplings, as we are confronted by the small value
of $\lambda_H$ in the standard model, and the apparent RG
behavior  $\lambda_H\rightarrow 0$ for scale of order $\sim 10^{12}$ GeV. For general gauge-Yukawa theories containing a scalar bilinear, the divergent behavior is readily obtainable \cite{Krog:2015bca}, but for the single doublet of the standard model, this is not easily constructed.

There is, however, a simple remedy available to us here: the Higgs portal interaction.
The main point is that if there exists new physics coupled
to the Higgs via a portal interaction, e.g., a sterile dark matter 
boson, then the Higgs Yukawa coupling that we observe,
$\lambda $ is actually only effective, and is replaced typically by a larger
value near the TeV scale  due to the mixing via the portal interaction \cite{general}.
We presently exploit this mechanism.

 We note that many authors have
considered various neutrino-composite Higgs boson schemes, many in the context of
a fourth generation and some with overlap to our present
case \cite{compnu}.
We turn presently to a Nambu--Jona-Lasinio schematic model of
our mechanism.

\section{NJL-Model}

The effective UV model we have in mind is a variation on
the Nambu--Jona-Lasinio model \cite{NJL} and 
top condensation models \cite{Nambu,Yamawaki,BHL}. We adapt this
to a neutrino condensate with the four-fermion interaction Lagrangian:
\beq
\label{zeroH}
\mathcal{L}'= \frac{g^2}{\Lambda^2} (\bar{L}_{Li}\nu^i_R)(\bar\nu_{Rj}{L}_{L}^{j})
+\frac{h{}^2}{\Lambda^2} (\bar{L}_{Li}\nu^i_R)(\bar{t}_{Ra}{T}_{L}^{a})+h.c.
\eeq
where   $L_L^i = (\nu^i,\ell^i)_L$ ($\nu_{Ri}$) are  left-handed lepton doublets (right-handed
neutrino singlets), 
and $T_L$ ($t_R$) is the top quark doublet (singlet); $(i,j,..)$ are generation indices
running to $N_f=3$ and $(a,b,..)$ are color indices
running to $N_c=3$.  The dominant large coupling
constant in our scheme is $g$ and $h < g/N_f$.  We will have additional
smaller couplings involving the other quarks associated with light fermion mass generation and flavor physics,
 as well as charge conjugated terms like $(\bar{L}_{Li}\nu^i_R)g^{jk}(\bar\nu_{Rj}{L}_{Lk})^C$.
These generate the charged lepton and quark masses and mixing angles, which we presently ignore. 

We follow \cite{BHL} and factorize the NJL interactions to write:
\begin{equation}
\label{oneH}
\mathcal{L}'_{\Lambda}=g \bar{L}_{iL} H \nu_R^i  + g' \bar{T}_{aL} H t^a_R -\Lambda^2 H^\dagger H
\end{equation}
Here we define $g' = h^2/g$.
Here
we have introduced an auxillary field $H$ that regenerates eq.(\ref{zeroH}) by $H$ equation
of motion.
This is the Lagrangian at the scale $\Lambda$, where the auxilliary field $H$ will
become the dynamical Higgs boson boundstate at lower energies.
We have ignored terms of order $g'^2$ which are generated when $H$ is integrated out to
recover eq.(\ref{zeroH}).  

We now use the RG to run the Lagrangian down to the Majorana mass scale, $M$, of the  
right-handed neutrinos, using only fermion loops.  The result is formally:
\bea
\label{twoH}
\mathcal{L}'_{M}& =& Z_H|DH|^2 -\widetilde{M}^2 H^\dagger H  + \frac{\widetilde{\lambda}_H}{2} (H^\dagger H)^2 
\nonumber \\
& &   \!\!\!\!\!   \!\!\!\!\!   \!\!\!\!\!   \!\!\!\!\! 
+[ g \bar{L}_{iL} H \nu_R^i  + g' \bar{T}_{aL} H t^a_R +\bar{\nu}^C_{Ri}M_{ij}\nu_R^j+ \text{h.c.}]
\eea
where the Majorana mass matrix, $M_{ij}$,  is now incorporated  by hand. 
The Higgs boson has acquired a logarithmic kinetic term and a quartic interacton
due to the fermion loops, and the Higgs mass has run quadratically:
\bea
Z_H & = & (4\pi)^{-2}( g^2N_f + g'{}^2N_c ) \ln( \Lambda^2/M^2) 
\nonumber \\
\widetilde{M}^2 & = & \Lambda^2 - (4\pi)^{-2}(2g^2N_f + 2g'{}^2N_c)(\Lambda^2 - M^2)
\nonumber \\
\widetilde{\lambda}_H & = & (4\pi)^{-2}(2g^4N_f + 2g'{}^4N_c)\ln({\Lambda^2}/M^2)
\eea
The quantities appearing in eq.(\ref{twoH}) are, of course, unrenormalized.
The renormalized couplings at the present level
of approximation are:
\beq
y_\nu = \frac{g}{\sqrt{Z_H}}\qquad y_t = \frac{g'}{\sqrt{Z_H}}\qquad \lambda_H = \frac{\widetilde{\lambda}_H}{Z^2_H}
\eeq
and we see that in the large $(N_f, N_c)$ limit the ratio $y^2_\nu/y^2_t$
is a constant. 

For simplicity, we take the Majorana mass matrix to be diagonal, $M=\text{diag} (M_1,M_2,M_3)$. 
In the large $M/v_{weak}$ limit, where $v_{weak}\sim 175$ GeV,
the masses of the three light neutrino states are given by the seesaw mechanism:
\begin{equation}
\label{numass}
m_{\nu}^i = \frac{y_\nu^2v_{weak}^2}{M_i},
\end{equation}
Assuming that $y_\nu$ is $\sim \mathcal{O}(1)$, and  $\sim eV$ masses for the light neutrinos, we expect $M_i~\sim~10^{13}$ GeV. Thus, in the  RG evolution of the system, loops containing right-handed neutrinos occur only above the scales
$M_i$.  As an approximation, take the threshold of the $\nu_R^i$ in loops to be at a common
Majorana mass scale $M$.

Note that the renormalized $\lambda_H = \widetilde{\lambda}_H/Z_H^2$ 
has the limits  $\lambda_H/y_\nu^4\rightarrow 0 $  and  $\lambda_H/y_\nu^2\rightarrow $~(constant)  as 
$\mu\rightarrow \Lambda$. 
The extent to which the top quark participates in the
binding of the Higgs boson relative to the neutrinos is determined by $g'{}^2N_c/g^2N_f$
which we assume is of order $1/N_f$, and thus the
 dominant coupling at the UV scale is $g^2$.
While we could keep the order $g'{}^2$ terms in the factorization of eq.(\ref{twoH}),  this
would make a weakly boundstate doublet, $H'$ composed mainly of $\bar{t}t$, but since $g'{}^2$ is   
subcritical this state would remain a heavy dormant doublet with $m^2 \sim \Lambda^2$. 

Below the Majorana mass scale $M$ the neutrinos decouple and the only significant
running in the fermion loop approximation is the top quark.  The electroweak scale is
tuned by the choice of  critical couplings.
The quadratic running to a zero mass Higgs boson, $\widetilde{M}^2=0$  defines
the critical coupling:
\beq
g^2N_f+ g'{}^2N_c=  8\pi^2 \left(  1+\frac{ M^2}{\Lambda^2}\right)
\eeq
The criticality, we assume, is due principally to the large value of $g^2$ and is only
slightly modified by the top quark. 
We then choose $g^2$ slightly supercritical
to produce the phenomenological tachyonic Higgs potential, 
 $\widetilde{M}^2=-M_H^2Z_H(\Lambda/M_H)$.

The NJL model is schematic, and must itself be an approximation to some new dynamics in
the UV.  This structure suggests a new gauge interaction which leads to eq.(\ref{zeroH})
upon Fierz rearrangement, in analogy to topcolor models \cite{topc},  as:
\beq
\label{zeroHnew}
\frac{g^2}{\Lambda^2} \bar{L}_{Li}\nu^i_R\bar\nu_{Rj}{L}_{L}^{j}
=
-\frac{g^2}{\Lambda^2} \bar{L}_{L}\frac{\lambda^A}{2}\gamma^\mu {L}_{L}\bar\nu_{R}\frac{\lambda^A}{2}\gamma_\mu\nu_R
+... \, ,
\eeq
where the Gell-Mann matrices $\lambda^A$ now act on the flavor indices.
The $g'$ term then requires some extension of the theory.
A model such as this assigns an $SU(3)$ gauge group to lepton family number,
and therefore gauge charges to the $\nu_{Ri}$, \ie, the $\nu_{Ri}$ are
no longer sterile.  This would imply that the Majorana
mass matrix must be generated by a VEV associated, \eg,  with additional $SU(3)$ scalar fields.
With  $\nu_{Ri}$ in the triplet represenation, this requires $\{ \bar{3} \}$ and/or $\{6 \}$
scalar condensates, and would dictate the neutrino mass and mixing angle structure. 
Construction of this kind of model will be done elsewhere.

\section{ Yukawa Sector}

The above discussion is the Wilsonian renormalization group approach.
To improve the calculation, we turn to the full RG equations which
are used below the scale $\Lambda$, together with the matching conditions dictated by the fermion bubble
approximation   \cite{BHL}. 
The  full RG equations (for $N_f=3$ these are a slight modification of ref.\cite{Martin}) take the form:
\bear
(4\pi)^2\frac{\beta_{y_t}}{y_t} &=& \frac{9}{2}y_t^2-8g_3^2+3\, \theta_M (\mu-M) y_\nu^2 -\frac{9}{4}g_2^2 -\frac{17}{12}g_1^2  \, , \nonumber\\
(4\pi)^2\frac{\beta_{y_\nu}}{y_\nu} &=& \theta_M(\mu-M)\left(\frac{9}{2}\, y_\nu^2+3y_t^2-\frac{9}{4}g_2^2-\frac{3}{4}g_1^2
\right),\nonumber\\
(4\pi)^2\frac{\beta_{g_1}}{g_1} &=& \frac{41}{6}g_1^2\, , \quad (4\pi)^2\frac{\beta_{g_2}}{g_2} = -\frac{19}{6}g_2^2\, \nonumber\\ 
\quad (4\pi)^2\frac{\beta_{g_3}}{g_3} & = & -7g_3^2 \, 
\eear
where $g_1,g_2$, and $g_3$ are the gauge couplings of the $U(1)_Y$, $SU(2)_L$, and $SU(3)_c$ symmetries respectively, $y_t$ is the top HY coupling, and $y_\nu$ the HY coupling of the lepton doublets to right-handed neutrinos. 
We have introduced a step-function,  $\theta_M=\theta(\mu-M)$, where
$\theta(x)=1$; $x \geq 0$ and $\theta(x)=0$; $x < 0$. The step-function models the threshold 
of the turn on of the right-handed neutrinos at the scale of the Majorana mass matrix.

In Fig.~\ref{yukawarun} we demonstrate the running of the HY coupling for the top quark and the neutrino as described. We use the initial conditions for the gauge couplings $g_1(m_Z)=0.36, \, g_2(m_Z)=0.65,\, g_3(m_Z)=1.16$, for the HY couplings: $y_t(m_t)=0.99$, and  $y_\nu(M)=1$, and for the masses $m_Z =91.2 \text{ GeV} ,\, m_t =173.2 \text{ GeV},$ and $ M=10^{13}  \text{ GeV}$.

\begin{figure}[b]
\begin{center}
\includegraphics[width=0.4\textwidth]{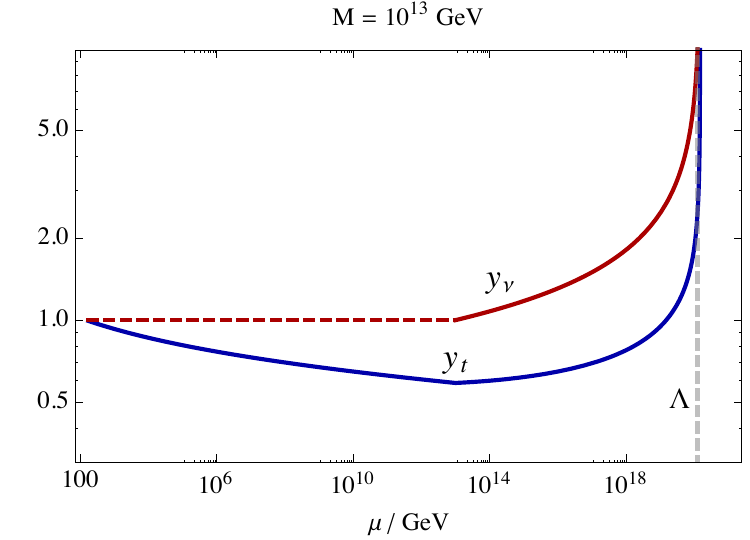}
\caption{The RG evolutions of the top (solid) and neutrino (dashed) HY couplings with contributions from right-handed neutrinos for renormalization scales above the neutrino Majorana mass; $\mu > M = 10^{13}$ GeV.}
\label{yukawarun}
\end{center}
\end{figure}
The evolution in Fig.~\ref{yukawarun} clearly indicates the existence of a Landau pole for the HY couplings at a scale $\Lambda \sim 10^{20} $ GeV, in accord with what one would expect if the Higgs is a fermion pair condensate. 

The Landau pole of the neutrino HY coupling  is seen to pull the top HY towards a Landau pole at $\Lambda$.
The neutrino HY coupling is always significantly larger than the top coupling for the scales where the perturbative result is valid. For the displayed example we find the ratio $y_\nu/y_t \geq 3$ for the region very close to the Landau pole.

To verify the consistency of this behavior of  
the top quark, consider the region below, but near, $\Lambda$.  Here the RG equations for
top and neutrino HY couplings can be approximated in the large $(N_f,N_c)$ limit by:
\beq
(4\pi)^2\frac{d\ln{y_t}}{d\ln\mu} \approx  
(4\pi)^2\frac{d \ln{y_\nu}}{d\ln\mu} \approx N_fy_\nu^2 + N_cy_t^2,  
\eeq
hence:
\beq
(4\pi)^2\frac{d\ln(y_t/y_\nu)}{d\ln\mu}=0
\eeq
This implies that  $y_\nu(\mu)/y_t(\mu) \rightarrow$ (constant),  as we approach the scale $\Lambda$.
The ratio $y_t/y_\nu \sim  g'/g$,
so the role of the top quark role is only that of a spectator. 

In this simplified setup, inserting an experimental neutrino mass in \eqref{numass} yields $y_\nu(M)$ as a function of $M$. For a chosen $M$, this value may be used as an initial condition in the RG equation for $y_{\nu}$, and the scale $\Lambda$ may be read off from the solution to the RG equations. A simple analytic estimate is given by setting to zero all couplings except $y_\nu$, in which case one finds for the one loop solution
\begin{equation}
\label{lambdaest}
\Lambda = M \,\text{exp}\left[\frac{(4\pi v_{weak})^2}{9 \, m_\nu^{exp}M}\right].
\end{equation}

Here $v$ is again the Higgs VEV and $m_\nu^{exp}$ is the experimentally measured neutrino mass.
The estimate \eqref{lambdaest} is in good agreement with the full numerical solution due to the fact that the neutrino coupling itself is what drives the divergence at $\Lambda$.
The relation \eqref{lambdaest} also gives a lower bound on the possible compositeness scale $\Lambda_{min}$ for any neutrino mass given by
\begin{equation}
\Lambda_{min} \simeq 1.5 \times \left(\frac{m_\nu^{exp}}{\text{eV}}\right)^{-1}\times10^{15}  \text{ GeV}.
\end{equation}

We perform the numerical analysis as before using the RG equations above, and obtain the scale associated with the Landau pole for different values of $M$ given a specific mass of the light neutrino states in the eV range. In Fig.~\ref{LvsM} we show numerical results concerning the relation between the Majorana mass and the $\Lambda$ scale for different values of the neutrino mass.
\begin{figure}[t]
\begin{center}
\includegraphics[width=0.4\textwidth]{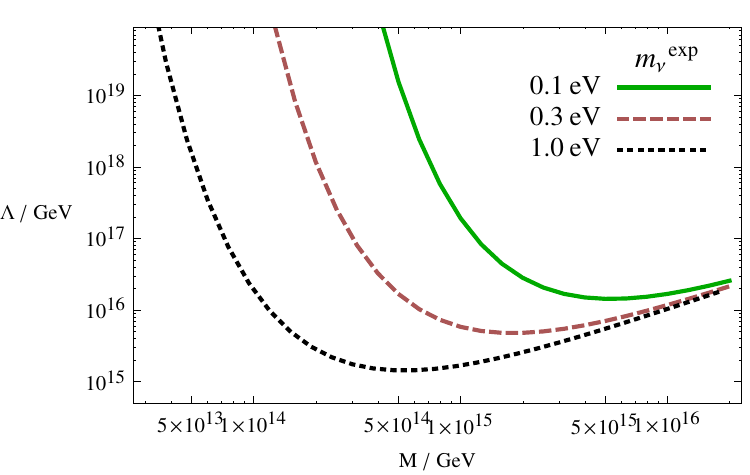} \hspace{.5cm}
\caption{Numerical results displaying the relation between the Majorana mass and the scale associated with the Landau pole for the neutrino HY coupling for different values of the neutrino mass.}
\label{LvsM}
\end{center}
\end{figure}
The perturbative nature of our analysis does not allow us to extrapolate to infinite coupling values, so we instead take the naive estimate of the $\Lambda$ scale to be defined by $y_\nu(\Lambda)=30$. We stress that this analysis is meant to provide a demonstration of principles rather than high precision results.

Two distinct behaviors are exhibited in Fig.~\ref{LvsM}: For smaller values of the Majorana mass, the scale $\Lambda$ is very sensitive to the choice of neutrino and Majorana mass. This is due to the fact that $y_{\nu}(m_t)$ is quite small for these values, and more RG time is needed to run to the Landau pole. For larger values of the masses, $y_{\nu}(m_t)$ also grows large in accordance with \eqref{numass} and the Landau pole is shifted closer to the scale where the neutrino coupling becomes active in the RG equations.

\section{scalar sector}

In the minimal version of a single composite Higgs boson, the physical Higgs mass prediction is larger than the  observed  $\sim 125$ GeV. 
The Higgs mass is controlled by the electroweak VEV, $v_{weak}$, and  the quartic coupling.  
The Higgs compositeness conditions predict a Landau pole for the quartic scalar coupling at the compositeness 
scale $\Lambda$ \cite{BHL}.  However, 
the quartic coupling constant in the standard model is to be too low to match these conditions, 
and indeed,  appears to decrease with scale potentially, becoming negative at $\sim 10^{12}$ GeV \cite{quartic}. 

To achieve compositeness of the Higgs boson, we
 employ a simple modification by which the observed Higgs quartic coupling, $\lambda$, becomes only a low energy {effective} coupling, while the true quartic coupling, $\lambda_H$, is larger and can have the requisite Landau pole.  The
actual quartic coupling needs only be about $2\times$ the observed $\lambda$ to achieve this, but requires additional
physics at the $\sim 1 $ TeV scale.

We extend the scalar sector to include a complex singlet \cite{general}, $S$ and the new Higgs potential becomes:
\begin{align}
V &=\frac{\lambda_H}{2} \left(H^\dag H  - v^2\right)^2 + \frac{\lambda_S}{2} \left(S^\dag S - u^2\right)^2 \nonumber \\ 
& \quad + \lambda_{HS}\left(H^\dag H  - v^2\right)  \left(S^\dag S - u^2\right),
\end{align}
where we have assigned the vacuum expectation values
\begin{equation}
\label{svevs}
\langle H^\dag H \rangle = v^2 ,\quad \langle S^\dag S \rangle = u^2. 
\end{equation}
The VEVs \eqref{svevs} are the global minima of the potential when $\lambda_H,\lambda_S > 0$ and $\lambda_H\lambda_S > \lambda_{HS}^2$. 

Expanding about the minimum of eq.\eqref{svevs}, one finds the mass matrix for the massive scalars to be
\begin{equation}
\frac{\partial^2V}{\partial \phi_i \phi_j}=
2
\begin{pmatrix}
 \lambda_H v^2 & \lambda_{HS} vu \\
\lambda_{HS} vu & \lambda_S u^2
\end{pmatrix}, \nonumber
\end{equation}
where $\phi_i$ refers to the direction of the vev in $H$ and $S$.

The eigenvalues are
\begin{equation}
m_\pm^2 = \lambda_H v^2 + \lambda_S u^2 \pm \kappa, \nonumber
\end{equation} 
where $\kappa= \sqrt{(\lambda_H v^2 -\lambda_S u^2)^2 + 4 \lambda_{HS}^2v^2u^2}$.
In the limit where $\lambda_H v^2 \ll \lambda_S u^2$, the lightest state mostly resides within $H$, and the mass can be approximated by
\begin{equation}
\label{mheff}
m_H^2 = m_{-}^2 = 2\left(\lambda_H - \frac{\lambda_{HS}^2}{\lambda_S} \right)v^2 + \mathcal{O}\left( \frac{\lambda_Hv^4}{\lambda_S u^2}\right) \, .
\end{equation}
The effective quartic coupling, measured from the Higgs mass, is now:
\begin{equation}
\lambda = \lambda_H - \frac{\lambda_{HS}^2}{\lambda_S},
\end{equation}
which is intrinsically smaller than the  coupling $\lambda_H$. Thus,  the composite picture 
with a suitable Landau pole in $\lambda_H$ is now possible.

\subsection{Singlet scalar extension}

We now analyze the RG evolution of the full theory with an eye to the landau
pole in $\Lambda_H$. Assuming  $S$ is an electroweak $SU(2)$  singlet, and $U(1)_Y$
sterile,
the RG equations for the scalar sector are given by
\bear
\beta_{\lambda_H} &=& (12 y_t^2+12\, \theta_My_\nu-3g_1^2-9g_2^2)\lambda_H - 12(y_t^4+\, \theta_My_\nu^4) \nonumber \\\ && + \frac{3}{4}g_1^4+\frac{3}{2}g_1^2g_2^2+\frac{9}{4}g_2^2+12\lambda_H^2+2\, \theta_u\lambda_{HS}^2, \\
\beta_{\lambda_{HS}} &=& \left(6 y_t^2+12\, \theta_My_\nu-\frac{3}{2}g_1^2-\frac{9}{2}g_2^2+6\lambda_H  \right) \lambda_{HS} \\ 
&& +4\, \theta_u( \lambda_S + \lambda_{HS} )\lambda_{HS} \, ,  \nonumber \\
\beta_{\lambda_{S}} &=& 4\lambda_{HS}^2 +10\, \theta_u\lambda_S^2,
\eear
where we have included the Heaviside function $\theta_u~=~\theta(\mu-u)$, to adjust for the fact that loops involving the $S$ state are not taking into account for scales below the vev $\langle S \rangle=u$ which generates the mass for the $S$ state. 

To accommodate the composite scenario, as first described in \cite{BHL}, both the quartic coupling and the HY coupling for the condensating fermions must diverge at a scale $\Lambda$. Furthermore, the nature with which the scalar becomes propagating at lower energy scales, sets the requirement
\begin{equation}
\underset{\mu \rightarrow \Lambda}{\text{lim}} \, \lambda_H /y_\nu^4 =0,
\end{equation}
and we expect the common divergence to yield
\begin{equation}
\label{lamyrat}
\underset{\mu \rightarrow \Lambda}{\text{lim}} \, \lambda_H /y_\nu^2 = \mathcal{O}(1).
\end{equation}

In Fig.~\ref{quarticflow} we demonstrate the evolution of the quartic coupling for a specific choice of initial conditions. We choose a mass for the active neutrino $m_\nu^{exp}=1$ eV, which yields a divergence of $y_\nu$ around $\Lambda = 10^{18}$ GeV, under the assumption that $M=5\times 10^{13}$ GeV. At the scale where $y_\nu(\mu)=10$, we then define the initial conditions for the quartic couplings $\lambda_H(\mu) = 98$, in accordance with \eqref{lamyrat}, and the somewhat arbitrary choices $\lambda_{HS}(\mu)=23$, $\lambda_S(\mu)=1.7$. The assumed value for $u=1$ TeV.
\begin{figure}[t]
\begin{center}
\includegraphics[width=0.4\textwidth]{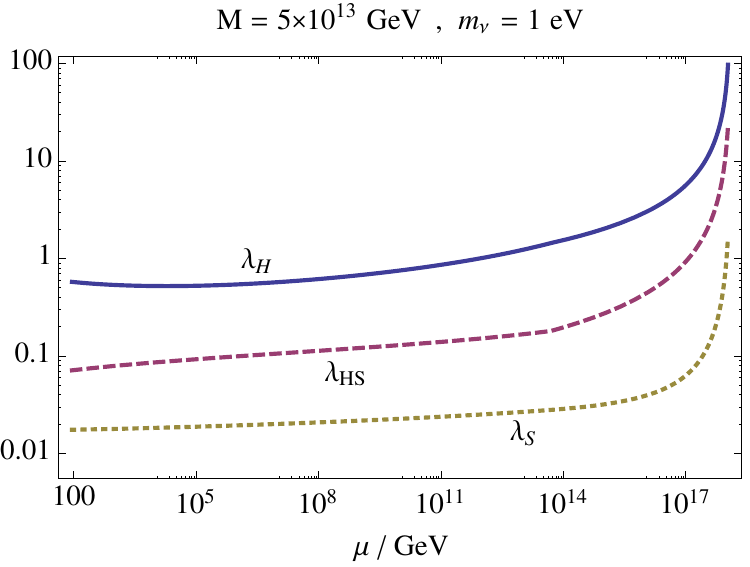} \hspace{.5cm}
\caption{The RG evolution for the quartic couplings for a specific choice of initial conditions at the UV scale. The IR phenomenology features a large quartic for the Higgs, while the effective coupling leads to a light Higgs mass $m_H\sim 130$ GeV.}
\label{quarticflow}
\end{center}
\end{figure}
The IR phenomenology features a large value for the Higgs quartic coupling $\lambda_H \sim 0.7$, while the effective coupling is considerably smaller $\lambda \sim 0.28$ corresponding to a Higgs mass $m_H \sim 130$ GeV.

The RG system involves some degree of tuning to ensure the proper behavior of the two new quartic couplings. Specifically, we must tune $\lambda_S$ to be small, to ensure a large correction in $\lambda$ as seen in \eqref{mheff} while $\lambda_{HS}$ is also tuned, such that $\lambda_H>\lambda_{HS}>\lambda_S$ is satisfied for all RG scales, in order to ensure a valid value of $\lambda$ at small scales. 

This model should merely serve as a proof of concept, displaying the possibility that the UV behavior of the Higgs quartic coupling can include a Landau pole. In this setup we have looked at the simplest possible scalar extension of the standard model with the standard Higgs mechanism in play for both scalars. The issues of tuned scalar couplings may then be alleviated if a different mechanism for symmetry breaking or a more complex scalar sector is considered. For a large class of more general gauge-Yukawa theories, a composite limit due to four fermion interactions at high energies is easily obtainable, as shown in \cite{Krog:2015bca}, while we will focus on the simplest alternative solutions below.

\section{Alternative scalar extensions}

As we have introduced a tuning between the dimensionless coupling constants of the scalar sector in addition to the usual tuning for the Higgs mass parameter, it would be beneficial to find a mechanism to stabilize the IR phenomenology towards changes in the initial UV conditions. We expect that this might be found by connecting the symmetry breaking mechanisms for the scalar sector.

In the previous example, the role of the ``portal" coupling $\lambda_{HS}$ was to supply a correction to the quartic Higgs coupling in the effective coupling by connecting the two scalar sectors, while the symmetry breaking mechanism is that of the standard Higgs boson for both the $H$ and $S$ scalars.
\subsection{Negative portal coupling}
We can expand the role of the portal coupling by letting the portal interaction communicate symmetry breaking in the dark $S$ sector to the standard model. Setting $\lambda_{HS}<0$ and assuming $\langle S \rangle \neq 0$ can trigger spontaneous symmetry breaking in the standard model, even if the mass term for the Higgs $m_H^2\geq 0$, since the portal interaction will add a negative squared mass contribution for $H$.
If the portal coupling is very small, then there can be a large hierarchy between the vevs of $S$ and $H$, and the validity of eq.\eqref{mheff} is guaranteed.

The change from a positive portal coupling to a negative one can thus change the nature of the symmetry breaking for the Higgs particle. It allows for other values of the Higgs mass parameter, and specifically one can choose $m_H^2=0$ and still obtain a second order phase transition due to the portal interaction. The actual analysis of this alternative model is however almost identical to the original, since the stability constraint and mass prediction only involves $\lambda_{HS}^2$. 
The measured Higgs mass is still obtainable together with a Landau pole for $\lambda_{H}$, albeit tuning between the scalar couplings is needed.

\subsection{Communicated CW symmetry breaking}

Common to the scalar sectors discussed so far has been the feature that a mass scale has been inserted by hand into the potential, either for both scalars, or for one of them. This enables the generation of a vast interval of possible scalar masses, but intrinsically means that these are very sensitive to the input parameters. An alternative way to generate mass scales is the dynamical one, where the mass scales arise directly from the RG evolution. We will show in the following that Landau poles in the quartic couplings, in accordance with a composite picture, may also accommodate spontaneous symmetry breaking due to the CW mechanism as demonstrated for elementary scalars in \cite{Hambye:2013dgv}.

It is central to the success of this model, that we now consider a dark\footnote{Similar models with a portal coupling to another scalar sector are often used to probe dark matter phenomenology.} scalar doublet $S$, gauged under a new $SU(2)_X$ group\footnote{The critical property of the gauge group is asymptotic freedom, so any other gauge group with this property could have been used.}. Since we want all mass scales to be generated dynamically, the potential is given as
\begin{equation}
V= \frac{\lambda_H}{2}(H^\dag H)^2 +\lambda_{HS}(H^\dag H)(S^\dag S)+\frac{\lambda_S}{2}(S^\dag S)^2,
\end{equation}
where we will investigate the cases where $\lambda_{HS}<0$.
Just as before, the requirement for stability of the potential is 
\begin{equation}
\label{stabcon}
\lambda_H>0, \quad \lambda_S > 0, \quad\lambda_H\lambda_S > \lambda_{HS}^2.
\end{equation}
Spontaneous symmetry breaking then occurs dynamically in this setup via the Coleman-Weinberg mechanism when the RG evolution brings the system of coupling constants into violation of the stability conditions \eqref{stabcon}. 

The driving force behind the symmetry breaking in this setup is the new gauge coupling $g_x$, related to the $SU(2)_X$ gauge symmetry. As this coupling becomes large at some scale due to asymptotic freedom, the quartic coupling $\lambda_S$ will be driven negative in the IR, due to the form of its beta function which is positive for any nonzero value of the couplings:
\begin{equation}
\label{betaS}
\beta_{\lambda_{S}} = 4\lambda_{HS}^2 +12 \lambda_S^2 + \frac{9}{4}g_x^4- 9g_x^2\lambda_S.
\end{equation}
 Denoting by $s^*$ the scale at which $\lambda_S = 0$, and performing the approximation close to this scale that $\lambda_S \simeq \beta_{\lambda_S} \text{ln} \left( \frac{s}{s^*}\right)$, the estimated value for the vev of $S$ coming from the associated Coleman-Weinberg symmetry breaking mechanism is given by
 \begin{equation}
 \label{svev}
 \langle S \rangle = u = s^* e^{-1/4}.
 \end{equation}
In return, the negative portal coupling $\lambda_{HS}$ induces a vev for $H$:
\begin{equation}
 \label{hvev}
\langle H \rangle = v = u \sqrt{\frac{-\lambda_{HS}}{\lambda_H}}.
\end{equation}
At this minimum, the mass matrix takes the form
\begin{equation}v^2
\begin{pmatrix}
2 \lambda_H  & -2\sqrt{-\lambda_H \lambda_{HS}}  \\
-2\sqrt{-\lambda_H \lambda_{HS}} & \lambda_{HS}-\beta_{\lambda_S}\frac{\lambda_H}{\lambda_{HS}} 
\end{pmatrix}.
\end{equation}
Assuming that $v^2 \ll u^2$ which is to say $\frac{-\lambda_{HS}}{\lambda_H} \ll 1$, we may expand the eigenvalues to the leading order in $\frac{\lambda_{HS}}{\lambda_H}$ and obtain
\begin{equation}
 \label{HSemasses}
m_1^2 = 2\lambda^2, \quad m_2^2 = -\frac{\beta_{\lambda_S}\lambda_H}{\lambda_{HS}}v^2,
\end{equation}
where the indices 1 and 2 relate to the state composed of mostly $H$ and $S$ respectively, and $\lambda = \lambda_H - \frac{\lambda_{HS}^2}{\beta_{\lambda_S}}$. This naturally resembles \eqref{mheff}, and we see once again, how the effective quartic coupling is smaller than the true coupling for the Higgs.

So far, the setup seems to resemble the simple one given in the previous chapter. The key difference is that while a high degree of tuning was needed for the initial conditions in the simple setup to guarantee the correct hierarchy at smaller scales, this is no longer the case, since the dynamics at these scales are controlled mainly by the evolution of the new gauge coupling.

Our probes of the parameter space for this theory will follow the lines of logic from the previous section: Assuming a certain neutrino mass $m_\nu$ and Majorana mass $M$, the scale of compositeness scale $\Lambda$ is determined uniquely. We will then impose the constraint \eqref{lamyrat}, which fixes the quartic couplings at this scale\footnote{We will assume that all quartic couplings are large at this scale which would be true in a theory where all scalars are composite in the sense we have described here. This is not a necessary assumption, and it may be relaxed if one wishes to consider elementary scalar dark matter extensions.}. The last remaining free parameter is the new gauge coupling $g_X$, which will be fixed at the mass of the Z boson. The only free parameters in our analysis is thus the two masses associated to the neutrino sector and the value $g_x(m_Z)$.

The RG equations for the remaining quartic coupling and the new gauge coupling is given to one loop by
\bear
 \nonumber\\
\beta_{\lambda_H} &=& (12 y_t^2+12\, \theta_My_\nu-3g_1^2-9g_2^2)\lambda_H - 12(y_t^4+\, \theta_My_\nu^4) \nonumber \\\ && + \frac{3}{4}g_1^4+\frac{3}{2}g_1^2g_2^2+\frac{9}{4}g_2^2+12\lambda_H^2+2\, \theta_u\lambda_{HS}^2, \\
\beta_{\lambda_{HS}} &=& \left(6 y_t^2+6\, \theta_My_\nu-\frac{3}{2}g_1^2-\frac{9}{2}g_2^2+6\lambda_H \right) \lambda_{HS} \\ 
&&  +\left(6 \lambda_S -\frac{9}{2}g_X^2\right) \lambda_{HS}   + 4\lambda_{HS}^2   \, , \nonumber \\
\beta_{g_{X}} &=& -\frac{43}{6}g_X^3 \, .
\eear
A numerical evaluation of the running of the couplings as described above will yield the vevs of $H$ and $S$ as well as the masses of the respective eigenstates, through \eqref{svev},\eqref{hvev}, and \eqref{HSemasses}, when the couplings are evaluated at the scale of symmetry breaking $s^*$.

A sample RG evolution yielding $v \simeq 175$ GeV and $m_H \simeq 125$ GeV is shown in Fig.~\ref{HSrun},
\begin{figure}[t]
\begin{center}
\includegraphics[width=0.4\textwidth]{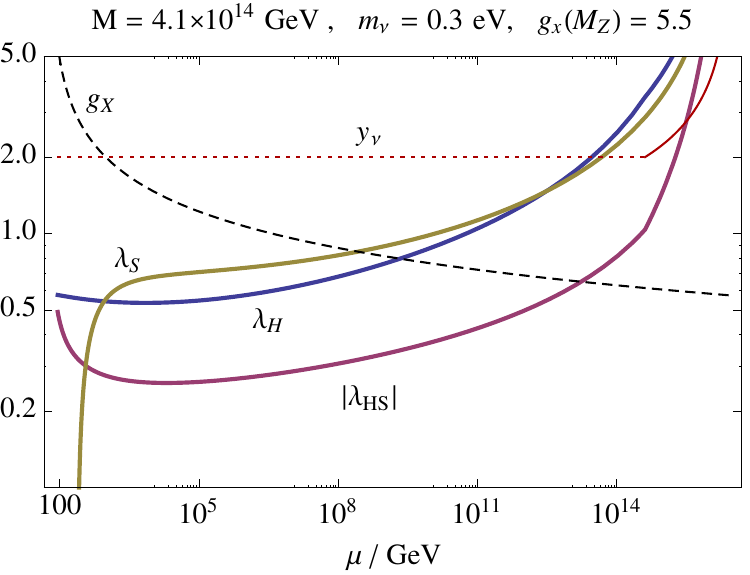} 
\caption{The RG evolution for the quartic couplings in the communicated CW setup. Choosing the active neutrino mass and Majorana mass, determines the compositeness scale, where the quartics are given values such that $\lambda_H = | \lambda_{HS} | = \lambda_S \approx y_\nu^2$ at this scale. The final assumption is that $g_X(m_Z) = 5.5$, which determines the IR behavior and symmetry breaking pattern. The evolution shown above yields $v \simeq 174$~GeV and $m_H \simeq 126$~GeV.}
\label{HSrun}
\end{center}
\end{figure}
where the increase of the gauge coupling $g_X$ in the IR is displayed alongside the decrease of the dark quartic $\lambda_S$, which is the source of the symmetry breaking. We warn the reader that the value for $\langle S \rangle = u \simeq 227$~GeV, such that $v^2/u^2 \sim 0.6$ such that the approximation used in \eqref{HSemasses} may be invalid and a more complete analysis should be performed. Once again, we postpone this for other work, while aiming for a qualitative description for now.

For the RG evolution shown above all quartic values are fixed to be equal at the compositeness scale, and the tuning between is no longer needed. Instead, having settled on a specific neutrino mass, only the Majorana mass $M$ and $g_X(m_Z)$ require balancing in order to get the correct phenomenology in the Higgs sector. Keeping $g_X(m_Z)$ fixed while varying $M$ with respect to the sample calculation above, yields the Higgs vev and mass depicted in Fig.~\ref{HiggsphenoM}.
\begin{figure}[t]
\begin{center}
\includegraphics[width=0.4\textwidth]{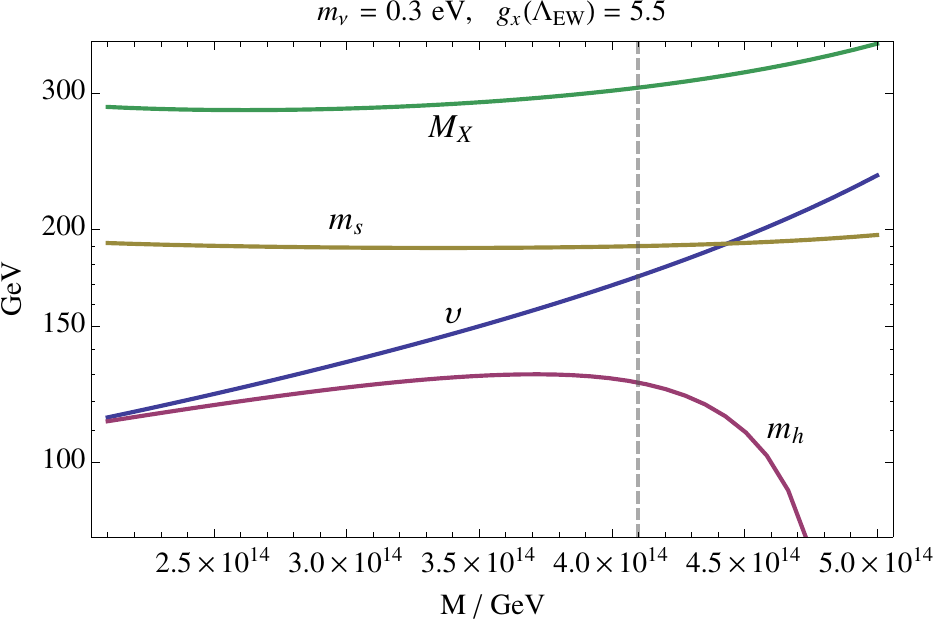} \hspace{.5cm}
\caption{Values of the Higgs vev ($v$) and mass ($m_h$) along with the mass of the extra scalar ($m_s$) and the dark gauge bosons ($M_X$) as the Majorana mass $M$ is varied, while the active neutrino mass $m_\nu = 0.3$ eV and $g_X(m_Z) =5.5$. The grey dashes indicate the point where correct Higgs phenomenology is realized.}
\label{HiggsphenoM}
\end{center}
\end{figure}
Interestingly, the Higgs mass seems to be stabilized around $\sim 130~\text{GeV}$ for a range of different Majorana masses, while the vev has a stronger dependence on $M$.

Varying $g_X(m_Z)$, one sees that in order to get values of $v$ and $m_H$ close to the correct values, one has to remain within the interval $g_X(m_Z) \in [5;6]$ with $M \sim 4 \times 10^{14}$~GeV for the chosen value of $m_\nu = 0.3$ eV. Thus the tuning problems within the parameters of the theory have been greatly reduced, and the interesting region of parameter space has been discovered. For the higher neutrino mass $m_\nu = 1$ eV, the relevant values of $M$ are centered at $M \sim 1.2\times 10^{14}$ GeV, while for the lower mass $m_\nu = 0.1$~eV, realistic Higgs phenomenology requires $M \sim 1.2 \times 10^{15}$~GeV, while the value of  $g_X(m_Z)$ is kept constant.

Along with the values for the Higgs mass and vev, we obtain values for the mass of the other scalar state $m_s = m_2$ from \eqref{HSemasses} along with the mass for the dark matter candidate $M_X = g_X*u/2$, which are also shown in Fig.~\ref{HiggsphenoM}. For the choice of parameters corresponding to the values for the Higgs observables, marked with a grey line, we obtain $m_S \sim 190$~GeV and $M_X\sim 300$~GeV. The predictions for these dark matter observables are fairly independent on the choice of neutrino and Majorana mass in the setup.

The phenomenology of the model presented here is by construction virtually identical to the one of its elementary counterpart, as reviewed in \cite{Hambye:2013dgv}. The main effect of imposing the composite picture is that the absolute value for the portal coupling $\lambda_{HS}$ is larger in our setup.

%
%
%
%

\section{Summary and Conclusions}

Our main goal in the present paper was to see how
difficult it is to maintain the idea  of a composite Higgs boson
in the sense of ref.\cite{Nambu,Yamawaki,BHL}, in light of
modern  standard model constraints.  While the
composite models
fine-tune the scale $v_{weak}$, they are in rough concordance with
the values of the Higgs boson and top quark masses as seen in nature,
and offer potential predictivity. 

 Nature appears at face value to resist the idea of a strong, dynamical fermion condensate
as the origin of the Higgs mechanism, given the apparent highly perturbative and critical behavior of the
quartic coupling $\lambda$.  It is, nonetheless, readily possible to construct a model
that can yield the compositeness conditions at large scale $\Lambda \sim 10^{15} -10^{19}$ GeV. 
Our main ingredient is the portal interaction that demotes $\lambda$ to
an effective low energy coupling, while the high energy theory is
controlled by $\lambda_H$.  We find a typical result that $\lambda_H \sim 2\times \lambda$.
This is sufficient to completely redefine the UV behavior of the theory. $\lambda_H$
can easily have a Landau pole and satisfy the Higgs boson compositeness
conditions \cite{BHL}.   Here we use a portal interaction near the TeV scale,
which is popular in a large number of scale invariant Higgs theories \cite{general}.

The constituents of the composite Higgs boson must couple with large Yukawa interactions to
the Higgs doublet, and these couplings must also have a Landau pole at
the scale $\Lambda$.
The top quark in the large $N_c$ fermion loop approximation in the standard model 
has too weak a Higgs Yukawa coupling to produce the Landau pole.  This is 
easier to solve than the $\lambda$ problem,  and one can imagine a
number of alternative theoretical fixes for it.  

Presently, however, we essentially abandon
the top quark as the constituent of the Higgs, and have followed
Martin \cite{Martin} to adopt the neutrinos as the Higgs boson constituents.
In the neutrino see-saw model \cite{seesaw} the Higgs will necessarily
have Yukawa couplings to the conventional left-handed 
lepton doublets, and right-handed neutrino singlets, $\sim y_\nu\bar{\psi}_L\nu_R\cdot H + h.c.$ .
These Yukawa couplings are not seen as $d=4$ operators in the low energy theory
below the scale $M$ of the right-handed neutrino masses, rather we observe only the $d=5 $ 
``Weinberg operators,''
$\sim y_\nu^2 (\bar{\psi}_L H)^c H{\psi}_L/M + h.c.$.  

Above the scale of the Majorana mass, $M$, the $d=4$ operators materialize and 
the Yukawa couplings, $y_\nu$, begin
to run.  We assume that neutrino mixing is driven by $M$, and 
assume degeneracy of three large Higgs Yukawas, $y^i_\nu$. Thus
our Higgs boson is engineered in a ``large $N_{flavor}=3$ fermion bubble approximation.''
The $y_\nu$ have a Landau pole and can match to the compositeness
conditions for the Higgs. The top quark Yukawa is also
pulled up to the Landau pole, but remains a spectator to the new dynamics
that forms the Higgs boundstate.  

The model has some nice features, lending a physical role to the right-handed 
neutrinos and demanding some new strong dynamics at $\Lambda$ (\eg, a gauged $SU(3)$ neutrino flavor?).
New dynamics near the weak scale is relevant for this.
As a proof of principle, there remains much to do to survey viable schemes
and explore their phenomenological consequences.

\vskip 10 pt

\noindent
{\bf Acknowledgements} 

\vskip 5 pt

This work was done at Fermilab, operated by Fermi Research Alliance, 
LLC under Contract No. DE-AC02-07CH11359 with the United States Department of Energy. J.K is partially supported by the Danish National Research Foundation under grant number DNRF:90.

\end{document}